\documentclass[%
 reprint,
 amsmath,amssymb,
pra,
]{revtex4-1}

\usepackage{graphicx}
\usepackage{dcolumn}
\usepackage{bm}
\usepackage{float}
\usepackage{graphics,graphicx}
\usepackage{amsmath,mathrsfs,amsfonts,amssymb,amsthm}
\usepackage{color}
\usepackage{tabularx}
\usepackage{subfigure}
\usepackage{color}
\usepackage{siunitx}    

\graphicspath{{../eps/}}
\DeclareGraphicsExtensions{.eps}

\def\mathscr{\mathcal}

\begin{document}

\preprint{APS/123-QED}

\title{Mixing of Spin and Orbital Angular Momenta via Second-harmonic Generation in Plasmonic and Dielectric Chiral Nanostructures}
\author{Xiaoyan~Y.Z.~Xiong$^1$}
\author{Ahmed Al-Jarro$^2$}
\author{Li~Jun~Jiang$^1$}
\email{ljiang@eee.hku.hk}
\author{Nicolae C. Panoiu$^2$}
\author{Wei~E.I.~Sha$^1$}

\affiliation{$^1$Department of Electrical and Electronic Engineering,~The University of Hong Kong, Hong Kong}%
\affiliation{$^2$Department of Electronic and Electrical Engineering, University College London, Torrington Place, London WC1E 7JE, United Kingdom}%

\begin{abstract}
We present a theoretical study of the characteristics of the nonlinear spin-orbital angular
momentum coupling induced by second-harmonic generation in plasmonic and dielectric nanostructures
made of centrosymmetric materials. In particular, the connection between the phase singularities
and polarization helicities in the longitudinal components of the fundamental and second-harmonic
optical fields and the scatterer symmetry properties are discussed. By in-depth comparison between
the interaction of structured optical beams with plasmonic and dielectric nanostructures, we have
found that all-dielectric and plasmonic nanostructures that exhibit magnetic and electric
resonances have comparable second-harmonic conversion efficiency. In addition, mechanisms for
second-harmonic enhancement for single and chiral clusters of scatterers are unveiled and the
relationships between the content of optical angular momentum of the incident optical beams and
the enhancement of nonlinear light scattering is discussed. In particular, we formulate a general
angular momenta conservation law for the nonlinear spin-orbital angular momentum interaction,
which includes the \emph{quasi-angular-momentum} of chiral structures with different-order
rotational symmetry. As a key conclusion of our study relevant to nanophotonics, we argue that
all-dielectric nanostructures provide a more suitable platform to investigate experimentally the
nonlinear interaction between spin and orbital angular momenta, as compared to plasmonic ones,
chiefly due to their narrower resonance peaks, lower intrinsic losses, and higher sustainable
optical power.
\end{abstract}

\pacs{42.25.-p, 42.65.Ky, 02.70.Pt}
\maketitle


\section{\label{sec:intro}Introduction}
Optical vortex beams carrying orbital and spin angular momentum have attracted great interest
because of their important technological applications, including to optical tweezers
\cite{oam1,oam2,oam3}, high-capacity optical communications \cite{oam4,oam5}, and quantum key
distribution \cite{oam6}. They have also been explored in various nonlinear optical processes
recently, e.g., orbital angular momentum (OAM) entanglement between photons created by spontaneous
parametric down conversion \cite{oam7}, OAM conservation in nonlinear wave mixing \cite{oam8}, and
enhanced nonlinear optical activity and nonlinear phase control \cite{czm16am,lcp15nm}. In this
context, the process of the second-harmonic (SH) generation (SHG), where two photons with fundamental frequency (FF), $\omega$, convert into one photon with SH frequency, $\Omega=2\omega$, is
the nonlinear process most used in studies of OAM, chiefly due to the relative simplicity of its
experimental implementation. Most previous studies in this area focused on the manipulation and
control of OAM by isotropic or chiral bulk nonlinear optical media during the SH conversion
process \cite{oam9,oam10,oam11,oam12,oam13}.

With recent advances in nanotechnology, it has become possible to investigate the SHG from
plasmonic and all-dielectric nanoparticles, which in most cases of practical interest are made of
centrosymmetric optical materials \cite{oam14,cpo07prb,bp10prb,oam15,bp13prl,oam16}. The SHG technique has
established itself as a powerful and promising tool for surface characterization, biomedicine, and
nanotechnology applications \cite{oam17,oam18,oam19,bp11n}. Despite this, a complete understanding
of the spin-orbit interaction (SOI) pertaining to the surface SHG in nanostructures is still
missing.

Plasmonic nanostructures that support localized surface plasmon-polaritons (SPPs) find pervasive
use in nonlinear optical devices, chiefly because plasmon-enhanced optical near-fields allow
nonlinear optical processes that ordinarily are weak to be significantly enhanced \cite{oam20}. On
the other hand, large ohmic losses in plasmonic materials limit the SH conversion efficiency and
optical power at which nonlinear plasmonic devices can be operated. Alternatively, dielectric
nanostructures exhibit a much lower intrinsic losses and hence a much higher power ablation
threshold, the drawback being that the local field enhancement is smaller than in the case of
plasmonic structures. Whereas a definitive conclusion has not been reached yet, preliminary
studies suggest that dielectric nanostructures might provide higher frequency conversion
efficiency as compared to their plasmonic counterparts, especially if resonant modes are employed
\cite{oam21,oam22}. In this connection, it should be stressed that investigations aiming to assess
the relative benefits provided by nonlinear plasmonic and dielectric nanostructures are
particularly helpful in clarifying this matter.

In this work, SOI in plasmonic and dielectric nanostructures upon the SHG process is studied by
using Laguerre-Gaussian (LG) beams~\cite{abp16oe}. These are the ubiquitous laboratory implementations of optical
vortices \cite{oam23,oam24}, their physical properties making them particularly suitable to study
the angular momentum of optical fields. In particular, LG beams in the LG$_{pl}$ configuration
have a concentric ring-like structure with $p$ radial nodes in the intensity profile and a
characteristic azimuthal phase variation of $e^{il\phi}$, where $\phi$ is the azimuthal coordinate
and $l$ is the topological charge of the beam measuring the phase variation, modulo $2\pi$, along
a closed curve containing the center of the vortex ($l$ is also called OAM photon number).
Generally, optical beams can also contain spin angular momentum (SAM) associated with their
polarization states, $\sigma$. Thus, $\sigma=0$ for linearly polarized light and $\sigma=1$
($\sigma=-1$) for left (right) circularly polarized (LCP, RCP) light. In this article, we provide
a comprehensive description of the scattering of circularly polarized LG beams from single
plasmonic and dielectric nanospheres made of centrosymmetric materials, chiral clusters made of
such nanospheres and with specific rotational symmetry properties, and chiral nanostructures with
identical symmetry properties as those of the nanosphere clusters. More specifically, we
investigate the characteristics of the SOI mediated by the considered nanostructures and its
relation to the properties of the linear and nonlinear (SHG) scattering processes.

The paper is organized as follows. In the next section we present the main results pertaining to
the linear and nonlinear light scattering from single nanoparticles as well as chiral clusters of
nanoparticles and chiral nanostructures. The SOI phenomena in these structures are analyzed by
exploring the properties of the near- and far-field. A general selection rule including the
\emph{quasi-angular-momentum}  matching for the chiral structures is also formulated. Furthermore,
we compare the SH conversion efficiency of chiral structures made of plasmonic and dielectric
materials. Finally, in the last section, we summarize the main conclusions of our study.

\section{\label{sec:results}Results and Discussions}
\begin{figure}[!b]
\begin{center}
\includegraphics[width=\linewidth]{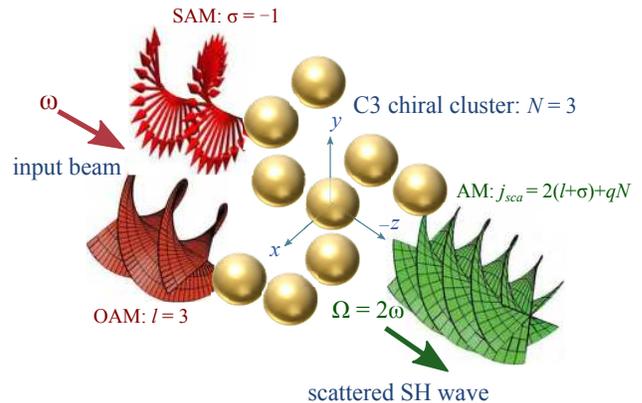}
\caption{Schematic of the nonlinear spin-orbital angular momentum (AM) coupling induced by second-harmonic generation (SHG) in a chiral cluster
with $N$-fold rotation symmetry. Both the frequency and angular momentum have doubled during the SHG process. In addition, the quasi-angular-momentum, defined as the order of the rotational symmetry of the chiral cluster ($N$), can be transferred to the scattered SH wave to conserve the total angular momentum. $q$ is an integer and $q=0,\pm1,\pm2,\pm3,\dots$.} \label{fig1}
\end{center}
\end{figure}

We consider a chiral cluster of nanospheres with $N$-fold rotation symmetry, which is surrounded by air
and irradiated by a circularly polarized LG beam carrying spin and orbital angular momentum, as shown in Fig~\ref{fig1}.
We calculate numerically the light scattering at the fundamental frequency (FF) and SH from
plasmonic and dielectric nanostructures using the boundary element method \cite{oamXiong}. Only the surface contribution to the SHG is considered for centrosymmetric materials~\cite{surfaceZhang, surfaceButet, surfaceHeinz}. The linear and nonlinear fields and currents at the surface of the nanostructure are used to characterize the SHG.

In this work we consider generic plasmonic and dielectric controsymmetric materials, namely gold
(Au) and silicon (Si). According to a model that is widely used in the study of the SHG in
centrosymmetric media, \cite{h91bookCh} the surface nonlinear polarization is described by a
surface nonlinear susceptibility and can be written as:
\begin{equation}\label{eq:nlsp}
    \mathbf{P}_{s}(\mathbf{r},\Omega) = \epsilon_{0}\hat{\chi}_{s}^{(2)}:\mathbf{E}(\mathbf{r},\omega)\mathbf{E}(\mathbf{r},\omega)\delta(\mathbf{r}-\mathbf{r}_{s}),
\end{equation}
where $\mathbf{r}_{s}$ defines the surface, $\hat{\chi}_{s}^{(2)}$ is the surface second-order
susceptibility tensor, $\mathbf{E}(\mathbf{r},\omega)$ is the fundamental field taken inside the material, and the Dirac function defines the surface characteristic of the nonlinear
polarization.

\begin{figure}[!b]
\begin{center}
\includegraphics[width=\linewidth]{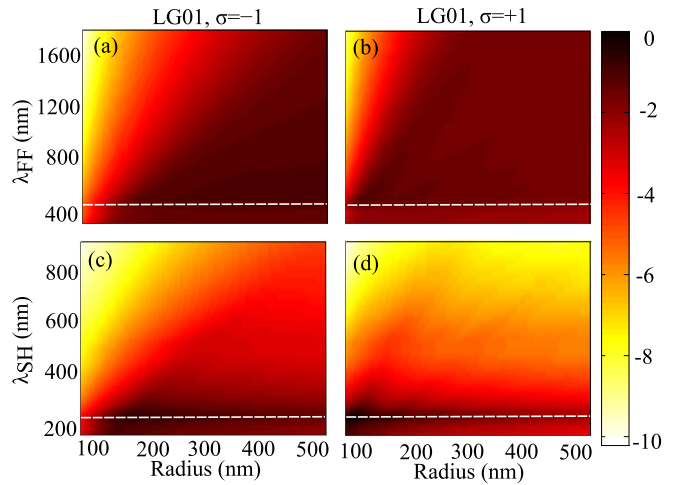}
\caption{Spectra of the normalized fundamental (top panels) and SH (bottom panels) scattering
cross section vs. radius of the gold sphere illuminated by LCP ($\sigma=+1$) and RCP ($\sigma=-1$)
LG beams with the OAM of $l=1$. The LG beam waist is $w_0=2\lambda_{\textrm{FF}}$. Dashed lines
correspond to the resonant responses in the bulk. Figures are plotted on a logarithmic scale.} \label{fig2}
\end{center}
\end{figure}

\begin{figure*}[t]
\begin{center}
\includegraphics[width=5.2in]{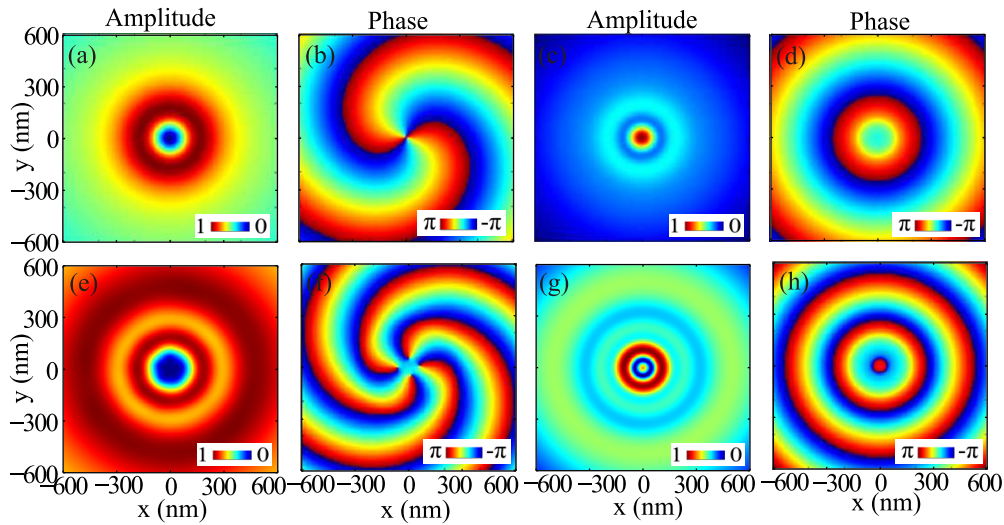}
\caption{Amplitudes and phases of the near-field longitudinal-field component ($E_z$) of the fundamental (top panels)
and SH field (bottom panels) scattered from a gold sphere illuminated by the LCP ($\sigma=+1$) and
RCP ($\sigma=-1$) LG beams with the OAM of $l=1$. The LG beam waist is
$w_0=2\lambda_{\textrm{FF}}$ and the wavelength of the beam is
$\lambda_{\textrm{FF}}=\SI{520}{\nano\meter}$. Radius of the gold sphere is $100\,\textrm{nm}$.
Panels (a), (b), (e), and (f) correspond to LCP beam ($\sigma=+1$), whereas panels (c), (d), (g),
and (h) to RCP beam ($\sigma=-1$).} \label{fig3}
\end{center}
\end{figure*}

Except for the case when the surface contains structural features with intrinsic chirality, the
surface of centrosymmetric media possesses an isotropic mirror-symmetry plane perpendicular to the
interface. Then, the surface nonlinear susceptibility $\hat{\chi}_{s}^{(2)}$ has only three
independent components, i.e., $\hat{\chi}_{s,\perp\perp\perp}^{(2)}$,
$\hat{\chi}_{s,\perp\parallel\parallel}^{(2)}$, and
$\hat{\chi}_{s,\parallel\perp\parallel}^{(2)}=\hat{\chi}_{s,\parallel\parallel\perp}^{(2)}$, where
the symbols $\perp$ and $\parallel$ refer to the directions normal and tangent to the surface,
respectively. The susceptibility components of Au are
$\hat{\chi}_{s,\perp\perp\perp}^{(2)}=1.59\times10^{-18}\,\textrm{m}^2/V$,
$\hat{\chi}_{s,\parallel\parallel\perp}^{(2)}=\hat{\chi}_{s,\parallel\perp\parallel}^{(2)}=4.63\times10^{-20}\,\textrm{m}^2/V$,
and $\hat{\chi}_{s,\perp\parallel\parallel}^{(2)}=0$ \cite{oam32}, while for Si they are
$\hat{\chi}_{s,\perp\perp\perp}^{(2)}=6.5\times10^{-18}\,\textrm{m}^2/V$,
$\hat{\chi}_{s,\parallel\parallel\perp}^{(2)}=\hat{\chi}_{s,\parallel\perp\parallel}^{(2)}=3.5\times10^{-19}\,\textrm{m}^2/V$,
and $\hat{\chi}_{s,\perp\parallel\parallel}^{(2)}=1.3\times10^{-19}\,\textrm{m}^2/V$ \cite{oam34}.

\begin{figure*}[t]
\begin{center}
\includegraphics[width=6.2in]{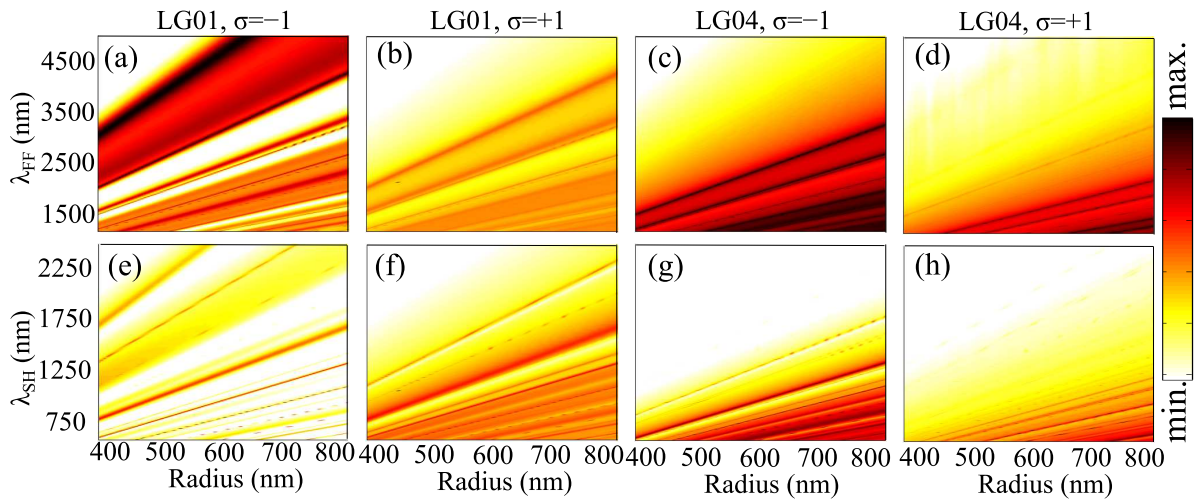}
\caption{Spectra of the normalized fundamental (top panels) and SH (bottom panels) scattering
cross section vs. radius of the silicon sphere illuminated by LCP ($\sigma=+1$) and RCP
($\sigma=-1$) LG beams with OAM of $l=1$ and $l=4$. The waist of the LG beam is
$w_0=2\lambda_{\textrm{FF}}$. Figures are plotted on a logarithmic scale.} \label{fig4}
\end{center}
\end{figure*}
\subsection{A Single Nanosphere}
We begin our study with an analysis of the scattering of polarized LG beams from a single gold
nanosphere, as the nanosphere possesses well-defined multipole-type optical modes. The gold
nanosphere placed at the system origin and embedded in vacuum is illuminated by a LG beam
propagating along the $-z$-axis with its minimum waist in the $xy$-plane. The LG beam carries a
topological charge, $l$, and it has no nodes along the radial co-ordinate (however, it can be zero at $r=0$), i.e. we
consider LG$_{0l}$ beams. In addition, we assume that the beam is circularly polarized, which
meaning it also carries SAM, i.e. $\sigma=+1$ for LCP and $\sigma=-1$ for RCP. The total angular
momentum carried by the circularly polarized LG beam is characterized by the integer $j=l+\sigma$.
In our analysis, the dielectric constant for gold is taken from experimental data \cite{oam31}.

The fundamental and SH scattering cross sections, as a function of the radius of the gold
nanosphere and the wavelength at the FF, $\lambda_{\textrm{FF}}$, are shown in Fig.~\ref{fig2}.
Due to the excitation of a surface plasmon resonance at the incident wavelength
$\lambda_{\textrm{FF}}=\SI{520}{\nano\meter}$, the gold nanosphere has the largest scattering cross section
for both LCP and RCP LG beams with the topological charge of $l=1$. The resonant responses are
marked with dashed lines in Figs.~\ref{fig2}(a) and \ref{fig2}(b).
The SH scattering cross-section maps exhibit a resonant response
exactly at $\lambda_{\textrm{SH}}=\SI{260}{\nano\meter}$ because of the plasmon-induced optical
near-field enhancement.

Although the resonances occur at the same frequency for LG beams with different polarization, the
amplitude and phase of the near-field fundamental and SH longitudinal-field components
($E_z$) have different
features due to the SOI. Figures~\ref{fig3}(a)--\ref{fig3}(d) and
Figs.~\ref{fig3}(e)--\ref{fig3}(h) display the magnitude and phase of  longitudinal fundamental and
SH fields, respectively, calculated in a section parallel to the $xy$-plane at the resonant
wavelength $\lambda_{\textrm{FF}}=\SI{520}{\nano\meter}$
($\lambda_{\textrm{SH}}=\SI{260}{\nano\meter}$). The radius of the gold sphere is fixed to
$\SI{100}{\nano\meter}$ and the observation plane is located $\SI{40}{\nano\meter}$
away from the back side of the sphere (the side furthest from the source).

It can be seen that, for the LCP light ($\sigma=+1$) the amplitude of the scattered fundamental
and SH fields vanishes at the axis of the beam due to the phase singularities at this point [see
Figs.~\ref{fig3}(a) and \ref{fig3}(e)]. Moreover, the phase profile of the SH field
[Fig.~\ref{fig3}(f)] has twice as many nodes along a closed path that contains the origin as
compared to the profile of the fundamental field [Fig.~\ref{fig3}(b)]. This means that during the
photon conversion process not only has the frequency doubled, but also the photon angular momentum
has doubled its value. In other words, both the energy and the total angular momentum are
conserved in the nonlinear interaction of light with the nanosphere.
In contrast, in the case of the RCP LG$_{0l}$ beam, the amplitudes of the forward scattered fundamental
[Figs.~\ref{fig3}(c)] and SH fields [Fig.~\ref{fig3}(g)] are no longer zero at the beam axis due to the
linear and nonlinear SOI upon resonant light scattering from the Au nanosphere. That is,
the phase vorticity related to OAM ($l=1$) cancels the polarization helicity related to SAM
($\sigma=-1$) in the incident beam, producing at both the FF and SH a field whose phase is independent on the azimuthal angle
[Figs.~\ref{fig3}(d) and \ref{fig3}(h)] without phase singularities at the beam axis.
This means that the value of $j=l+\sigma$ is preserved
during the light interaction with the nanosphere, i.e. the total angular momentum is conserved.
This is not surprising as the sphere has $\mathcal{C}_{\infty}$ symmetry with respect to rotations
around the beam axis and consequently angular momentum cannot be transfred between the beam and
nanosphere.

\begin{figure}[t]
\begin{center}
\includegraphics[width=\linewidth]{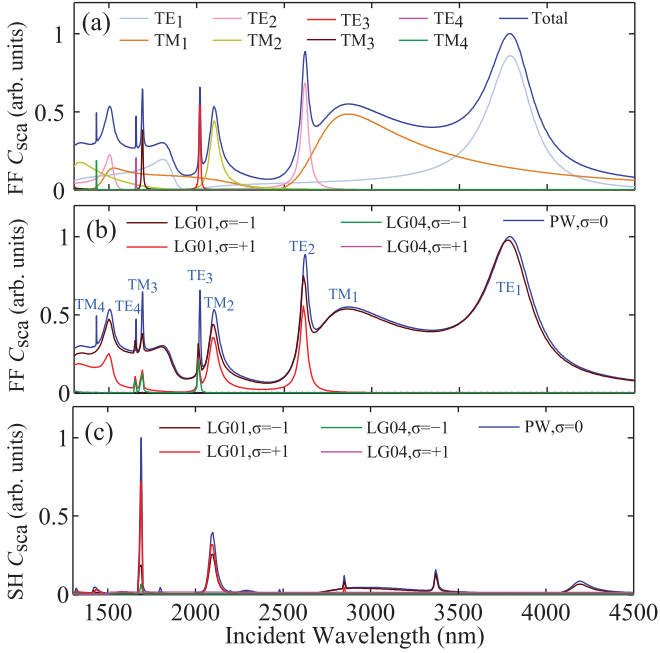}
\caption{Normalized scattering cross section, $C_{\textrm{sca}}$, spectra of a silicon nanosphere
with radius of $\SI{500}{\nano\meter}$. The incoming excitation is a linearly polarized plane wave
or circularly polarized LG beams. (a) Fundamental $C_{\textrm{sca}}$ and its multipole expansion
for linearly polarized plane wave excitation. (b) Fundamental $C_{\textrm{sca}}$ for linearly
polarized plane wave and circularly polarized LG beam excitations. (c) SH $C_{\textrm{sca}}$ for
linearly polarized plane wave and circularly polarized LG beam excitations.} \label{fig5}
\end{center}
\end{figure}

In order to understand the specific differences between the characteristics of scattering of LG
beams from metallic and dielectric nanoparticles, we considered next the scattering of such beams
from silicon nanospheres. The linear scattering process has been discussed in \cite{abp16oe}, but
for the sake of completeness we also summarize here the main ideas. The radius of the nanosphere
is chosen so that the wavelengths of the optical resonances are located in the spectral regions
where the intrinsic optical losses of silicon can be neglected, i.e.
$\lambda\gtrsim\SI{1.2}{\micro\meter}$, and the wavelength dependence of the permittivity of
silicon is described \textit{via} the Sellmeier equation~\cite{oam33}. The incident LG beams carry
an OAM of $l=1,4$ and SAM of either $\sigma=+1$ or $\sigma=-1$. Figure~\ref{fig4} shows the
fundamental and SH scattering cross section maps as a function of the nanosphere radius and
wavelength. Different from the broad-spectrum optical response from a single gold sphere (see
Fig.~\ref{fig2}), the spectra of the silicon nanosphere are much richer in spectral features, with
a series of resonant bands, which become narrower and more closely spaced as the wavelength
decreases. The richer spectra of the silicon nanosphere is due to two effects:
one is the size effect, that is the larger silicon nanosphere supports more resonances; the other is
the excitation of magnetic modes in the silicon nanosphere as will be illustrated later.
Equally important, with the increase of the total angular momentum $j=l+\sigma$, more
resonant bands are suppressed in the long-wavelength region for both the fundamental
[Figs.~\ref{fig4}(a)--\ref{fig4}(d)] and SH [Figs.~\ref{fig4}(e)--\ref{fig4}(h)] fields. As will
be explained later, this phenomenon is closely related to the amount of angular momentum carried
by the LG beams and that contained in the optical (Mie) modes of the silicon nanosphere.

The physical origin of these resonance bands can be understood by considering a simpler scattering
configuration, i.e. a linearly polarized plane wave ($\sigma=0$). Thus, we calculate the
scattering spectrum of a single silicon nanosphere with the radius of $\SI{500}{\nano\meter}$
excited by a linearly polarized plane wave. Figure~\ref{fig5}(a) illustrates the decomposition of
the scattering cross section in terms of transverse magnetic (TM) and transverse electric (TE)
multipoles, as the silicon nanosphere supports both electric and magnetic optical modes. When
comparing the scattering spectra corresponding to illumination with plane wave and LG beams with
$l=1$ and $l=4$, as displayed in Fig.~\ref{fig5}(b), we found that the resonance peaks of the LG
beams are located at the same positions as those of the plane wave excitation. More specifically,
the scattering spectrum of the RCP LG beam with $l=1$ and $\sigma=-1$ is almost identical to that
corresponding to the linearly polarized plane wave. This is because the incident field profile of
the LG beam is similar to that of the plane wave, and that the latter is achiral,
considering that the polarization helicity and
phase vorticity cancel each other and the beam waist $w_0=2\lambda_{\textrm{FF}}$ is much larger
than the radius of the silicon nanosphere.

and that the latter is achiral

Because of the conservation of the projection along the beam axis of the total angular momentum,
the LG$_{0l}$ beam with $\sigma=\pm1$ can only excite multipole modes of the nanosphere with $n\ge
l+\sigma$ ($n$ is the angular momentum quantum number of spherical
harmonics $Y_{nm}$, $m$ is the magnetic quantum number, $-n\le m \le n$ and $n\ge 1$),
and therefore all multipole modes with radial quantum number smaller than
$l+\sigma$ will be completely suppressed. This is illustrated in Fig.~\ref{fig5}(b), where the
LG$_{01}$ beam with $\sigma=+1$ only excites modes with $n\ge2$, i.e. the TE$_1$ and TM$_1$
resonances vanish, whereas the LG$_{04}$ beam with $\sigma=-1$ only excites modes with $n\ge3$,
the scattering channels of TE$_1$, TM$_1$, TE$_2$, and TM$_2$ being completely suppressed.
Similarly, the LG$_{04}$ beam with $\sigma=+1$ only excites optical modes with $n\ge5$.

Figure~\ref{fig5}(c) presents the SH scattering cross section spectra. When comparing
Figs.~\ref{fig5}(b) and \ref{fig5}(c), we can observe that both the fundamental and SH spectra
contain resonances at several specific wavelengths of the incident beam, e.g. at
\SI{1691}{\nano\meter}, \SI{2100}{\nano\meter}, and \SI{2865}{\nano\meter}. The strong SH
radiation at these wavelengths is due to the resonant enhancement of the optical field at the FF,
that is when fundamental field resonance occurs at these wavelengths, the local field is enhanced
and thus more SH light is generated. However, we can also find in the SH spectra resonances that
have no counterpart in the linear spectra, e.g. those at
$\lambda_{\textrm{FF}}=\SI{3382}{\nano\meter}$ and $\lambda_{\textrm{FF}}=\SI{4200}{\nano\meter}$.
These peaks are due to the fact that optical modes exist at half the wavelength of the excitation
beam. For example, when the fundamental wavelength is \SI{3382}{\nano\meter}, SH currents are
generated at \SI{1691}{\nano\meter}. These SH sources resonantly excite the optical mode of the
sphere that exists at this frequency and, consequently, the scattered SH field is significantly
enhanced.
\begin{figure}[t]
\begin{center}
\includegraphics[width=\linewidth]{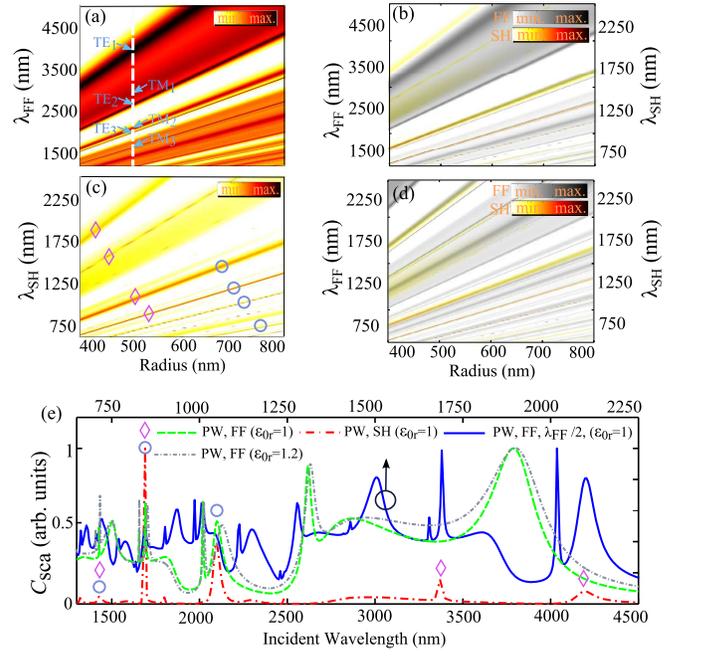}
\caption{Spectra of the normalized scattering cross section vs. radius of the silicon sphere
illuminated by linearly polarized plane waves. (a) fundamental field and (c) SH field with incident
wavelength ranging from \SIrange{1300}{4800}{\nano\meter}; (b) overlap between maps (a) and (c);
(d) overlap between map (c) and the map at the FF [$\lambda_{\mathrm{FF}}$ ranges from \SIrange{650}{2400}{\nano\meter}];
(e) fundamental and SH scattering cross section of the silicon sphere with radius fixed to \SI{500}{\nano\meter} surrounded by background material with different relative permittivity $\epsilon_{0r}$ ($\epsilon_{0r}=1$ and $\epsilon_{0r}=1.2$).
In (b,d), the left ($\lambda_{\mathrm{FF}}$) and right ($\lambda_{\mathrm{SH}}$) axes are used for the fundamental
and SH scattering cross section, respectively. Figures (a)-(d) are plotted in a logarithmic scale. Figure (e) is plotted in a linear scale.}
\label{fig6}
\end{center}
\end{figure}

In order to further clarify the origin of these two kinds of SH enhancement mechanisms, we compare
in Figs.~\ref{fig6}(a) and \ref{fig6}(c) the maps of the fundamental and SH scattering cross
sections corresponding to linearly polarized plane wave excitation ($\sigma=0$), the incident
wavelength ranging from \SIrange{1300}{4800}{\nano\meter}. Figure~\ref{fig6}(b) shows the overlap
between the map at the FF [$\lambda_{\mathrm{FF}}$ ranges from \SIrange{1300}{4800}{\nano\meter},
Fig.~\ref{fig6}(a)] and SH map [$\lambda_{\mathrm{SH}}$ ranges from
\SIrange{650}{2400}{\nano\meter}, Fig.~\ref{fig6}(c)]. The overlapping resonance bands, denoted by
circles in Fig.~\ref{fig6}(c), demonstrate the FF-resonance enhancement mechanism. Moreover, we
also calculated the map of the scattering cross section at the FF with the excitation wavelength
ranging from \SIrange{650}{2400}{\nano\meter}. The overlap between this map and the map of the
scattering cross section at SH, with $\lambda_{\mathrm{SH}}$ ranging from
\SIrange{650}{2400}{\nano\meter}, is shown in Fig.~\ref{fig6}(d). The overlapping bands correspond
in this case to the SH-resonance enhancement mechanism  and are denoted by diamonds in
Fig.~\ref{fig6}(c). Interestingly enough, there are two double-resonance bands exhibiting both the
FF-resonance and SH-resonance enhancement mechanisms, the corresponding intensity of the generated
SH being particularly large when this double-resonance condition is satisfied.

\begin{figure}[b]
\begin{center}
\includegraphics[width=\linewidth]{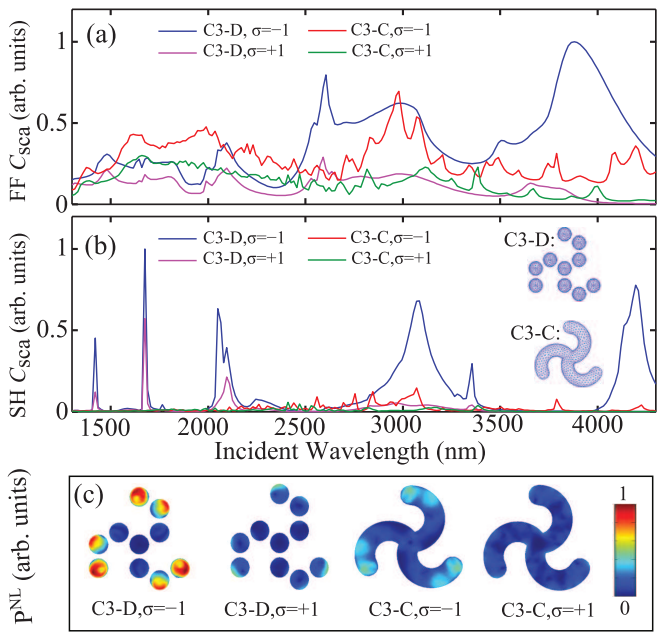}
\caption{Spectra of the normalized scattering cross section, $C_{\textrm{sca}}$, of dielectric
chiral nanostructures illuminated by LCP ($\sigma=+1$) and RCP ($\sigma=-1$) LG beams with the OAM
quantum number of $l=4$. The LG beam waist is $w_0=2\lambda_{\textrm{FF}}$. (a) $C_{\textrm{sca}}$
at the FF. (b) $C_{\textrm{sca}}$ at SH. (c) Spatial distributions of the nonlinear polarization
currents at the surface.} \label{fig7}
\end{center}
\end{figure}
\subsection{Chiral Nanostructures with Rotational Symmetry}
Chiral nanostructures have been widely investigated for generating or interacting with OAM beams~\cite{oam38Williams, oam39Coles, oam40Sakai}. Armed with the knowledge pertaining to the linear and nonlinear light scattering from a single
nanosphere, we consider now the scattering of LG beams from planar chiral nanostructures~\cite{oam41Menzel} with $N$-fold
rotational symmetry. For the sake of specificity, we set $N = 3$ in our calculations. To begin
with, we consider a chiral nanostructure consisting of 10 identical nanospheres arranged in an
Archimedes-like spiral configuration, as schematically shown in the inset of Fig.~\ref{fig7}(b),
and denoted as \textsf{C3-D}. The second chiral nanostructure is the ``continuous'' counterpart of
the first one and is denoted as \textsf{C3-C} in the inset of Fig.~\ref{fig7}(b). The chiral
nanostructures lie in the $xy$-plane in such a way that their center coincides with the origin of
the coordinate system. The LCP and RCP LG beams with the OAM of $l=4$, LG$_{04}$, are assumed to
propagate along the $-z$-direction with the beam axis passing through the origin. Moreover, we
assume that in both cases the chiral nanostructures are made of silicon because in this case the
spectra of the scattering cross-section have a larger set of features. For a more reliable
comparison, the radius of the spheres in the cluster is assumed to be equal to the radius of the
cross section of the arms of the continuous \textsf{C3-C} nanostructure and set to
\SI{500}{\nano\meter}.

The scattering spectra at the FF were calculated and are depicted in Fig.~\ref{fig7}(a).
Despite the ``noisy" aspect of spectra in Fig.~\ref{fig7}, important features can be deduced. It shows
that the scattering characteristics of the chiral structures strongly depend on the polarization
state of the LG beam. Thus, for the LG$_{04}$ beam with the RCP ($\sigma=-1$), the quantum number
of the total angular momentum of the incident beam is $j=l+\sigma=3$, which matches the rotational
symmetry order of the two chiral structures, $N=3$. This \emph{quasi-angular-momentum} matching
between the incident beam and the chiral scatterers greatly increases the light scattering at the
FF. Although the \textsf{C3-C} nanostructure illuminated by the RCP LG$_{04}$ beam shows strong
light scattering in the linear regime, this is not the case for its SH radiation, which is
presented in Fig.~\ref{fig7}(b). Comparing the optical response of the two structures, one can see
that the \textsf{C3-D} nanostructure has larger radiation intensity for both the fundamental and
SH fields.

These results clearly prove the critical role played by the resonances of single nanospheres in
the scattering processes at the FF and SH. In particular, the scattering spectra of the
\textsf{C3-D} nanostructure have clearly defined and pronounced spectral peaks corresponding to
resonances supported by individual nanospheres in the cluster. To better illustrate the SH
enhancement stemming from excitation of nanosphere resonances, we plot in Fig.~\ref{fig7}(c), for
both chiral nanostructures, the distribution of the surface nonlinear polarization currents. It is
evident that the \textsf{C3-D} cluster illuminated by the LG$_{04}$ beam with $\sigma=-1$ displays
the largest nonlinear polarization currents, which in turn give rise to the strongest SH
radiation.
\begin{figure}[t]
\begin{center}
\includegraphics[width=\linewidth]{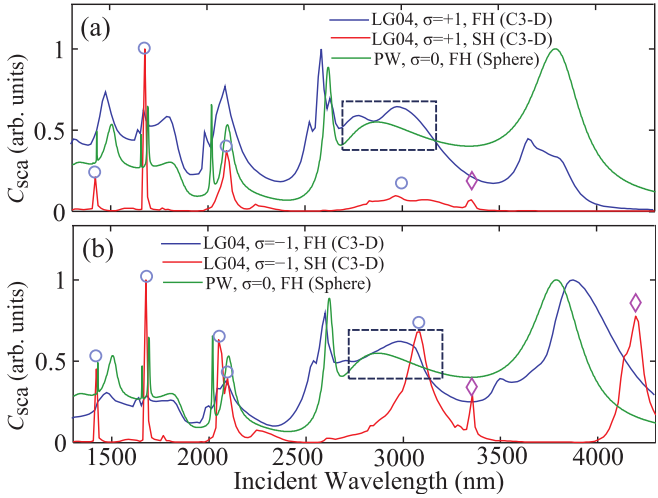}
\caption{Comparisons between the fundamental and SH scattering spectra of the silicon chiral
cluster with $3$-fold rotational symmetry and the fundamental scattering spectra of a single
silicon nanosphere. (a) LG$_{04}$, $\sigma=+1$. (b) LG$_{04}$, $\sigma=-1$.} \label{fig8}
\end{center}
\end{figure}

In order to reveal the physical origin of the SH resonant peaks for the chiral nanostructures, we
compare the fundamental and SH scattering spectra of the silicon chiral cluster and the
fundamental scattering spectra of a single silicon nanosphere, as per Fig.~\ref{fig8}. The
incident waves are the LG beams and plane waves for the chiral cluster and the signal nanosphere,
respectively. At the FF, the spectral locations of the scattering peaks of the chiral cluster
almost coincide to those corresponding to a single nanosphere, except for small deviations at long
wavelengths caused by the mutual optical coupling between the nanospheres. The SH peaks around
\SI{3000}{\nano\meter} show a broadband enhancement resulting from the mutual coupling, which also
can be confirmed by the resonance splitting [Fig.~\ref{fig8}(a)] and shift [Fig.~\ref{fig8}(b)]
marked by dashed boxes in
the fundamental spectra of the cluster. On the other hand, high-order whispering gallery modes,
which are modes whose OAM is different from zero, supported by a dielectric sphere are more
confined and have lower radiation loss at shorter wavelengths, leading to weaker coupling. This
suggests that the strong fundamental scattering of the chiral cluster is primarily due to the
resonances of individual nanospheres in the cluster. The SH scattering peaks having both FF and
SH resonances enhancements are marked in Fig.~\ref{fig8} by circles and diamonds,
respectively. We stress that the resonance peaks located in the long wavelength region are due to
the SH-resonance enhancement mechanism \textit{in the cluster}, because the resonance scattering
peaks of a single nanosphere are suppressed at long wavelengths, as illustrated in
Fig.~\ref{fig5}.

\begin{figure}[t]
\begin{center}
\includegraphics[width=\linewidth]{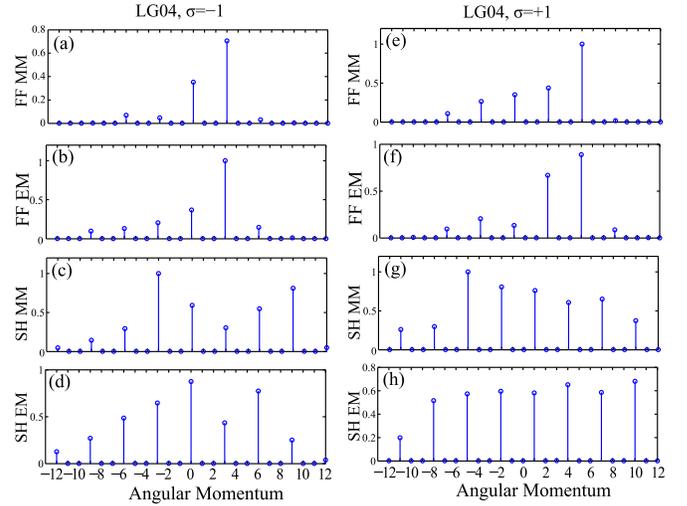}
\caption{Multipolar decompositions of the fundamental and SH scattered waves from the silicon
\textsf{C3-D} chiral cluster in terms of angular momentum channels. MM and EM denote magnetic
multipole and electric multipole, respectively. The silicon \textsf{C3-D} chiral cluster is
illuminated by the RCP LG$_{04}$ beam (left panels) and LCP LG$_{04}$ beam (right panels). (a),
(b), (e), and (f) correspond to scattered waves at FF, whereas (c), (d), (g), and (h) correspond
to wave scattering at SH.} \label{fig9}
\end{center}
\end{figure}
For the nanostructures with $N$-fold rotational symmetry, the incident beam with well-defined
angular momentum can only scatter into a set of modes with specific values of the total angular
momentum. In order to describe how nanostructures with $N$-fold rotational symmetry affect the
transfer of angular momentum from the incident field to the scattered field during both linear and
nonlinear interaction processes, we formulate a general angular momentum conservation law. It can
be written as:
\begin{eqnarray}\label{eqR1}
j_{sca}=s(l+\sigma)+q N,
\end{eqnarray}
where $l$ and $\sigma$ are the orbital and spin angular momentum numbers of the incident beam,
respectively, $s$ denotes the order of harmonic generation ($s=1$ for linear processes,
$s=2$ for
SH generation, etc.), $N$ is the quasi-angular-momentum number possessed by the nanostructure with
$N$-fold rotational symmetry, $q$ is an integer $q=0,\pm1,\pm2,\pm3,\dots$,
and $j_{sca}$ is the total angular momentum
number of the scattered field. Equation~\eqref{eqR1} implies that, the total angular momenta of
the scattered fundamental or SH fields depend not only on the angular momenta of the incident beam
but also on the rotational symmetry properties of the scatterer. The quasi-angular-momentum of the
scatterer can be transferred to the scattered wave in order to conserve the total angular
momentum. Note that this is the rotational analogue of the conservation of momentum upon
scattering of electromagnetic waves diffraction gratings, when the quasi-momentum of the grating
must be taken into account when describing the light scattering from such optical structures.

To validate this angular momentum conservation law, the total angular momentum of the scattered
fundamental and SH fields from the \textsf{C3-D} chiral cluster are analyzed by the multipole
expansion method \cite{Jackson}. Figure~\ref{fig9} shows the relative contributions of the
electric and magnetic multipoles to the total angular momentum of the scattered fields. For the
incident LG beam with $l=4$ and $\sigma=-1$, the quantum numbers characterizing the total optical
angular momentum are $j_{sca}=0,\pm3,\pm6,\pm9,\dots$, for both the scattered fundamental and SH
fields. For the incident LG beam with $l=4$ and $\sigma=+1$, the scattered fundamental field has
the quantum numbers $j_{sca}=\dots,-7,-4,-1,2,5,8,\dots$, whereas the scattered SH field has the
quantum numbers $j_{sca}=\dots,-11,-8,-5,-2,1,4,7,10,\dots$. These predictions, based on
Eq.~\eqref{eqR1}, completely agree with the multipole decomposition weights of the total angular
momentum of the scattered fundamental and SH fields presented in Fig.~\ref{fig9}.

\begin{figure}[t]
\begin{center}
\includegraphics[width=\linewidth]{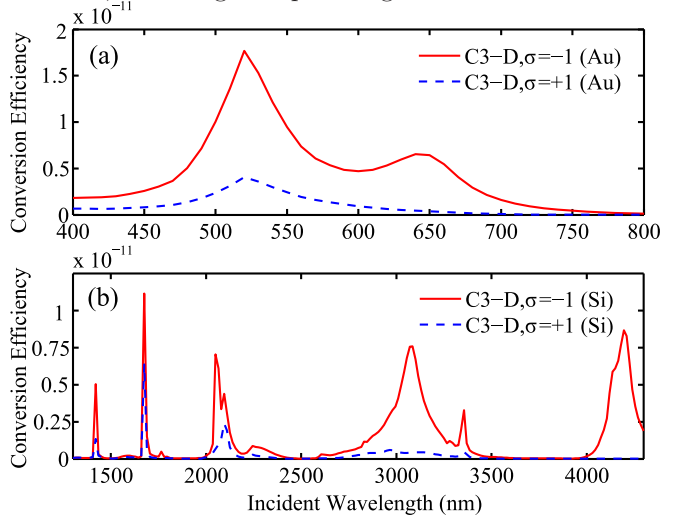}
\caption{SH conversion efficiencies of \textsf{C3-D} chiral clusters illuminated by LCP and RCP
LG$_{04}$ beams. (a) Plasmonic (gold) chiral cluster. (b) Dielectric (silicon) chiral cluster.}
\label{fig10}
\end{center}
\end{figure}

The scattering of circularly polarized LG$_{04}$ beams from plasmonic chiral clusters is also
investigated. The plasmonic chiral cluster has the same configuration as the silicon one and is
made of gold. The radius of each gold nanosphere is chosen to be \SI{100}{\nano\meter} so that the
cluster as a whole has strong surface plasmon resonances in the optical frequency region. The SH
conversion efficiencies of the plasmonic (gold) and dielectric (silicon) chiral clusters
illuminated by LCP and RCP LG$_{04}$ beams are compared in Fig.~\ref{fig10}. Note that the
conversion efficiency of the dielectric chiral cluster is normalized to the cross-section area of
the plasmonic and dielectric nanosphere, respectively, so that it gives us a per-particle
quantification of the nonlinear conversion efficiency. The power of the incident LG beam for the
dielectric and plasmonic nanostructures is set to be same, $P_{\mathrm{in}}=2\,\textrm{W}$. As we
can see in Fig.~\ref{fig10}, the dielectric chiral cluster made of high-permittivity silicon
nanospheres supports both electric and magnetic resonances, leading to similar conversion
efficiency when compared to that of the plasmonic chiral cluster.
Considering that the much lower optical losses in dielectrics
allow them to sustain higher optical power, it is possible that the dielectric
nanostructures can provide orders of magnitude higher SH conversion efficiency, as
compared to their plasmonic counterparts.

\section{Conclusion}
In summary, we have studied the scattering of circularly polarized Laguerre-Gaussian beams from
single plasmonic and dielectric nanospheres made of centrosymmetric materials, as well as chiral
clusters made of such nanospheres, by means of the boundary element method. The phase
singularities and vorticities in the longitudinal components of the fundamental and
second-harmonic fields have been discussed and their relationship with the rotational symmetry of
the scatterer has been analyzed. The doubling of the angular momentum and frequency during the
second-harmonic nonlinear spin-orbit-interaction process has been demonstrated. The resonant
excitation of multipolar optical modes in the dielectric nanosphere, leading to significant
enhancement of scattered radiation at second-harmonic, has been analyzed in detail. A general
angular momenta conservation law, including the quasi-angular-momentum of chiral nanostructures
with $N$-fold rotational symmetry, has been formulated to describe how the rotational symmetry
properties affect the transfer of optical angular momentum from incident beams to both the
scattered fundamental and second-harmonic fields. Our study reveals that, as compared to plasmonic
nanostructures, dielectric nanostructures with pronounced resonance peaks and high ablation
optical power threshold are better candidates for applications in near-field optical microscopy,
nonlinear biosensing, optical angular momentum multiplexing/demultiplexing, and quantum
information coding and decoding.

\section{Acknowledgements}
This work was supported by the Research Grants Council of Hong Kong (GRF 716713, GRF 17207114, and GRF 17210815), National Science Foundation of China (Nos. 61271158 and 61201122), Hong Kong ITP/045/14LP, Hong Kong UGC AoE/-04/08, the Collaborative Research Fund (No. C7045-14E) from the Research Grants Council of Hong Kong, and Grant CAS14601 from CAS-Croucher Funding Scheme for Joint Laboratories. This project is supported in part by a Hong Kong UGC Special Equipment Grant (SEG HKU09). N.C.P. acknowledges support from the European Research Council, Grant Agreement no. ERC-2014-CoG-648328. The first author thanks Prof. W.C.H. Choy and Prof. W.C. Chew for constructive discussions and suggestions.


\end{document}